\documentclass[a4paper,twocolumn,11pt,accepted=2025-12-02]{quantumarticle}
\pdfoutput=1
\usepackage[utf8]{inputenc}
\usepackage[english]{babel}
\usepackage[T1]{fontenc}
\usepackage{amsmath}
\usepackage{hyperref}
\PassOptionsToPackage{compress}{natbib}

\usepackage[numbers,sort&compress]{natbib}

\usepackage{graphicx}

\usepackage{braket}
\usepackage{lipsum}
\usepackage{bm}
\usepackage{dsfont}
\usepackage{amssymb}
\usepackage[dvipsnames]{xcolor}
\usepackage{pstricks}
\usepackage{verbatim}
\usepackage{units}
\usepackage{mathtools}
\usepackage{multirow}
\usepackage{enumitem}
\usepackage{mathrsfs}
\usepackage{leftidx}
\usepackage{xspace}
\usepackage{amsmath}
\usepackage[normalem]{ulem} 
\usepackage{soul}
\usepackage{physics}
\usepackage{tabularx}
\usepackage{upgreek}
\usepackage{cancel}

\newcommand{\oPi}{\overline{\Pi}}
\newcommand{\e}{{\rm e}}

\usepackage{hyperref}
\hypersetup{colorlinks=true,linktoc=all,linkcolor=blue,breaklinks=true,citecolor=blue,urlcolor=blue}

\usepackage[english]{babel}
\begin{document}

\title{Purely quantum 
  memory in closed systems observed via imperfect measurements}

\author{Jorge Tabanera-Bravo}
\email{jtabane@mpinat.mpg.de}
\author{Alja\v{z} Godec}
\email{agodec@mpinat.mpg.de}
\affiliation{Mathematical bioPhysics group, Max Planck Institute for Multidisciplinary Sciences, Göttingen 37077, Germany}

\begin{abstract}
    The detection and quantification of non-Markovianity, a.k.a.\ memory, in quantum
    systems is a central problem in the theory of open quantum
    systems. There memory is a result of the interaction between the
    system and its
environment. Little is known, however, about memory effects
induced by imperfect measurements on closed systems, where an entanglement with the environment is not possible. We investigate the emergence and
characteristics of memory in closed systems observed via
imperfect stroboscopic quantum measurements yielding coarse-grained outcomes. We consider ideal and two kinds of imperfect measurements: von Neumann
 measurements---the analogue of classical lumping---which destroy any
 coherence in the system, and \emph{genuinely
 quantum-lumping}  L\"uders measurements preserving certain quantum
 correlations. Whereas the conditions for Markov dynamics under von
 Neumann lumping are
 the same as for
classical dynamics,
quantum-lumping requires stronger
conditions, i.e.\ the absence of any detectable coherence. We introduce the concept of \emph{purely
quantum memory} having no classical counterpart. We illustrate our results with a quantum walk on
a lattice and discuss their implications for dissipative dynamics and
decoherence effects induced by more realistic measurements. 
\end{abstract}

\maketitle

\section{Introduction}

In 1932 John von Neumann established that upon a quantum measurement,
the quantum state of a system is projected (or collapsed) onto a one-dimension
subspace of the Hilbert space associated with the measurement
outcome~\cite{von2018mathematical}. In stark contrast to classical
measurements, quantum measurements have drastic effects on the system's
evolution because they destroy the quantum
coherence, giving rise, e.g.\ to the Zeno effect
\cite{gardiner2004quantum} and quantum localization
\cite{percival1998quantum}. In the case when quantum measurements
are performed stroboscopically and one may also consistently define a
first-detection time (as the quantum analog of the first-passage time)
\cite{Meidan2019Aug,Yin2019Nov, Thiel2018Jun, Friedman2017Mar}, 
the destruction of coherence has been
associated with the ``classicality'' and the Markov property of the
quantum evolution \cite{Strasberg2019Aug, Smirne2018Nov, Milz2020Dec}.

The
situation becomes much more intriguing if measurement outcomes are
degenerate; according to von Neumann the collapsed
state should then correspond to a \textit{statistical mixture} of one-dimensional
projections. However, in 1950 Gerhart L\"uders noted that these
projections are (i)
 \emph{not} uniquely defined for a degenerate measurement, and (ii) a
 degenerate measurement preserves some quantum
coherence \cite{Luders1950Jan,Luders2006Sep}. 
Nowadays the Lüders concept of measurement under ideal conditions
\cite{Hegerfeldt2012Mar} is fully accepted, but the von Neumann
postulate still remains useful in several circumstances
\cite{Budroni2014Jul,Heinosaari2010Sep, SudheerKumar2016Oct,
  Footnote_1} where the measurement is affected by noise.
\begin{figure}[h!]
    \centering
    \includegraphics[width=\linewidth]{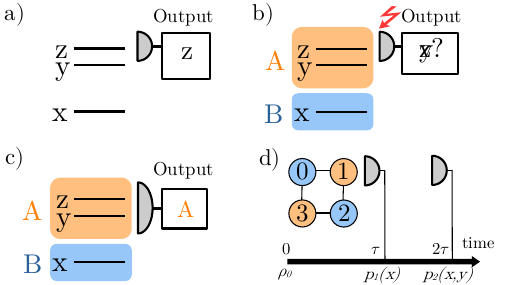}
    \caption{a) Sketch of an ideal quantum measurements:~The
      measurement apparatus distinguishes between all individual
      quantum states.~b) von Neumann measurement:~A quantum state is
      measured but the noise makes the outcome indistinguishable
      amongst groups of certain states. The latter become \emph{lumped}
into distinguishable sets called \emph{mesostates} $A = \{z,y\}$, $B =
\{x\}$..., which are clearly differentiated by the measurement
apparatus.~c) L\"uders measurement of a single mesostate:~The
measurement apparatus is intrinsically unable to distinguish sets of
states, such that the output are directly lumped mesostates.~d)
Stroboscopic measurement of a quantum system. The system shown is a quantum lattice with 4 sites and two mesostates (blue-orange). }
    \label{fig:esquema}
\end{figure}

We look at this old question from a conceptually different angle, i.e.\ from the
point of view of a  non-Markovian evolution on a \emph{lumped}
(coarse-grained) quantum state
space (see
\cite{Haenggi_1,Grigolini,Massi_CG,Rahav,Puglisi,Teza,Talkner,martinez,hartichCommentInferringBroken2022,David,Blom_2024,holography,Zhao,Tassilo}
for classical examples). Namely, both von Neumann's and L\"uder's
measurements
\textit{lump} indistinguishable individual outcomes onto
distinguishable sets called \textit{mesostates}. 
The dynamics projected onto mesostates preserves memory of the past
evolution, inducing non-Markovian statistics \cite{Haenggi_1,Grigolini,Lapolla2019Nov,
  Brandner2025Jan}. In the classical setting memory is known to emerge when
statistically distinct microscopic transitions
become projected onto mesostate-transitions as a  ``state-space
contraction''
\cite{Haenggi_1,Grigolini,Lapolla2019Nov,Brandner2025Jan,David,Blom_2024,holography,holography,Zhao,Tassilo}. The
detection, quantification, and understanding of quantum non-Markovian evolutions is
even more subtle and is a central problem in open quantum systems
theory \cite{breuer2002theory, Rivas2014Aug, Breuer2016Apr,
  Bernardes2017Mar, Banerjee2025Jul} inspiring numerous dedicated experiments
\cite{Li2011Jun,Smirne2011Sep,Cialdi2014Nov,Gessner2014Feb,Tang2015Dec}. \color{black}
In open quantum systems, memory is well known to be a result of the interaction between the system and its environment \cite{breuer2002theory, Rivas2014Aug, Breuer2016Apr, Bernardes2017Mar, Pollock2018Jan, Pollock2018Jan2, Taranto2019Apr, Taranto2019Apr2, Giarmatzi2021Apr, Taranto2024May}. However, since the quantum measurement process directly affects the system's dynamics, different measurement protocols imply starkly different memory effects \cite{Guo2021Jun}.

In this work, we describe how L\"uders and von-Neumann type
coarse-grained measurements give rise to different kinds of memory. In
particular, we find that the von-Neumann measurement exhibits memory
effects similar to a classical non-Markovian process, while the
L\"uders measurement gives rise to purely quantum features. The L\"uders
measurement requires stronger conditions to recover Markovian
behavior.\\\\The article is organized as follows: in Section~\ref{sec:Imperfect} we define the effects of ideal von-Neumann and
L\"uders measurements on a quantum system. In
Sec.~\ref{sec:consecutiv} we describe the evolution of quantum
systems under general \textit{consecutive} measurements, and discuss
how the different measurement schemes imply non-Markovian lumped
statistics. To conclude, in Sec.~\ref{sec:example} we present an
example of a real system where von-Neumann measurements give rise to
fully Markov statistics, whereas L\"uders measurements cause purely
quantum memory effects. 

\color{black}

\section{Imperfect measurements} \label{sec:Imperfect}

\color{black}

A prospective quantum measurement that can not determine completely
the state of the system, is called an \textit{imperfect measurement};
otherwise, the measurement is \textit{ideal}. The typical origin of
imperfect measurements is noise affecting the detector, inherently disabling
an accurate record of measurement outputs of a projective measurement \cite{Datta2011Jun, Len2022Nov}. Other examples include weak measurements \cite{Yin2019Nov,Yin2025Jun, Wang2024Oct, Rosset2012Dec}, non-demolishing \cite{Grangier1998Dec,   Eckert2008Jan, Johnson2010Sep, Kono2018Jun, Sewell2013Jul}, hidden
qubits \cite{Pechal2021Nov}, or indirect measurements
\cite{Thiel2018Jun, Pokorny2020Feb}. In these examples, the
information about the actual state of the system is coarse-grained (or
\textit{lumped}) into a set of fewer
measurement outcomes called
\textit{mesostates}. We observe that in such a noisy measurement, the
resulting quantum state is 
fully collapsed, i.e. the resulting
state is a classical mixture \cite{Len2022Nov}. However, in certain
measurements, such as the weak measurement, the resulting state does
\emph{not} fully collapse certain subspaces of the system. Following the
discussion in the Introduction, we call \textit{von-Neumann
  measurements} those that fully collapse the state of the system
during the process, and \textit{L\"uders measurements} those which
preserve certain quantum coherences.

In this section, we state the models of \textit{ideal} von-Neumann and
L\"uders measurements, respectively, depicted in
Fig.~\ref{fig:esquema}a-c. \color{black}
Let the possible measurement outcomes be $x,y,z$, e.g.\ the energy
levels of the hydrogen atom, or the sites of a quantum lattice. The
ideal measurement (Fig.~\ref{fig:esquema}a) distinguishes between
individual outcomes; the probability of obtaining each outcome $x$ is
given by the Born rule $p(x) = \Tr{\Pi_x\rho_0}$, where $\Pi_x =
\ket{x}\bra{x}$ is a rank-1 projector associated with the outcome $x$
and $\rho_0$ the density matrix representing the initial preparation
of the quantum system. After the measurement, the initial state is
collapsed onto the corresponding outcome subspace $\rho_0\rightarrow
\Pi_x\rho_0\Pi_x/p(x)$. 

Now consider the same measurement but performed by a noisy apparatus,
see Fig.~\ref{fig:esquema}b. Here the probability of each outcome is
still given by the Born rule. However, the thermal or electric noise
renders the exact readout impossible and hence produces blurred
outcomes $z,y...$, such that we can only resolve mesostates $X =
\{z,y,...\}$ these blurred outcomes belong (i.e.\ are
\emph{classically lumped}) to. 
The probability of obtaining an outcome inside mesostate $X=A$ is
$p^C(A) = p(z) + p(y) + ... = \sum_{x \in A} p(x)$ (analogously for
every mesostate). The quantum state of the system is fully collapsed,
such that the resulting state after the measurement is the weighted
sum of all the possible outcomes in $A$, \color{black}$\rho_0\rightarrow \rho_A^C =
\sum_{x \in A} \Pi_x\rho_0\Pi_x / p^C(A)$\color{black}. This corresponds to the von Neumann
measurement.

In the third setting (see Fig.~\ref{fig:esquema}c) the apparatus is
unable to distinguish $z,y$ even in the absence of noise. Here, the measurement outcomes are directly the
mesostates $X_i=A, B,...$. The probability of observing the mesostate $A$
is $p^Q(A) = \Tr{\oPi_A\rho_0\oPi_A}$ with the high-rank projector
$\oPi_A \equiv \Pi_z + \Pi_y + ... = \sum_{x\in A} \Pi_x$. After the
measurement the quantum state is projected according to $\Pi_A$, \color{black}
$\rho_0\rightarrow \rho_A^Q = \oPi_A\rho_0\oPi_A/p^Q(A)$\color{black}, corresponding to 
the L\"uders measurement. Notably, the resulting
$\rho_A^Q$ is a block diagonal matrix; the non-diagonal elements
represent the quantum coherence surviving the L\"{u}ders measurement,
preserving information about the previous state of the system. We will
show that this \emph{purely quantum} memory effect prevents Markovian
observed evolutions. 
\color{black} To distinguish the two, we will 
in the following refer to the statistics $p^C(A)$ as \textit{classical
  lumping} and  to $p^Q(A)$ as \textit{quantum lumping}.

We remark that the outcomes of the L\"uders and von-Neumann
measurements are identical, they only differ in the resulting quantum
state. 
A direct method to distinguish these measurements corresponds to
performing quantum tomography of the state after measurement
\cite{SudheerKumar2016Oct}.  However, one can also differentiate
between these measurement types by performing consecutive
measurements. For example, in \cite{Yin2019Nov, Yin2025Jun,
  Wang2024Oct} the authors consider the role of surviving coherence
along repeated L\"uders measurements on the statistics of the first detection-time.

\section{Quantum system under consecutive measurements} \label{sec:consecutiv}

Following the previous discussion we now consider the scenario of
consecutive measurements on a quantum system. In this scenario, the
quantum system is instantaneously measured at discrete times $t_1,
t_2, ...$. Between any two consecutive measurements, the system's evolution is given by an arbitrary trace-preserving map $\Lambda(t_1, t_2)$, $\rho(t_2) = \Lambda(t_1, t_2) \circ \rho(t_1)$. For example, if the system is isolated the evolution is given by the \textit{unitary} map $U(t_2-t_1) = \exp[-iH(t_2-t_1)/\hbar]$, $\rho(t_2) = \mathcal{U}(t_2-t_1) \circ \rho(t_1) = U\rho(t_1)U^\dag$, where $H$ is the Hamiltonian of the system. 

The joint probability distribution of the outcomes of $n+1$ consecutive measurements is \cite{gardiner2004quantum}
\begin{align}\label{eq:prob_definicion}
&  p[\{o_i,t_i\}_{i\in [0,n]}] = \nonumber \\
&  \qquad\Tr{P_{x_n}\circ\Lambda(t_n, t_{n-1})\circ P_{x_{n-1}}... \circ \rho_0},
\end{align}
were $P_{x_n}\equiv\Pi_{x_n}$ if the measurements are ideal with
outcomes $o_i \equiv x_i$, and $P_n \equiv \oPi_{X_n}$ if the
measurements are of the L\"uders type with $o_i \equiv X_i$. As we
show below, the statistics of von-Neumann outcomes is obtained by
lumping statistics of ideal measurements. 

The memory effects in Eq.~\eqref{eq:prob_definicion} can manifest in
several ways. In the conventional picture, the quantum system is in
contact with a certain environment, able to retain some information
about the past evolution of the system \cite{breuer2002theory,Alonso2005May, Piilo2008May,Rivas2014Aug,Breuer2016Apr,
  deVega2017Jan}. Due
to this memory, the maps $\Lambda(t_n, t_{n-1})$ could depend on 
previous measurement outcomes. Within this framework, the
non-Markovianity of the system is usually identified with the
divisibility of the maps $\Lambda(t_n, t_{n-1})$ into
completely-positive intermediate maps \cite{Rivas2014Aug}. However,
this approach can not capture the memory effects introduced by
different measurement instruments acting on the system 
\cite{Milz2019Jul}. There exist different approaches to Markovianity
trying to reconcile the memory effects of the environment with the
consecutive-operations framework \cite{Pollock2018Jan,
  Pollock2018Jan2, Taranto2019Apr, Taranto2019Apr2}. The common point
in any consistent definition of quantum Markovianity is recovering
the classical definition in the classical limit, this is, that
Eq.~\eqref{eq:prob_definicion} yields
\begin{equation}\label{eq:markodefinicion}
    \begin{split}
        p[x_n, t_n|x_{n-1}, t_{n-1}; x_{n-2}, t_{n-2};...] =  \\
        p[x_n, t_n|x_{n-1}, t_{n-1}].
    \end{split}  
\end{equation}
In any other case the process is non-Markovian. From this one can
conclude that there is a unique way of ``Markovian'', but there are
multiple fundamental ways of ``non-Markovian''. A natural distinction
amounts to identifying different environments with \textit{classical
  memory}, and \textit{quantum memory} (or entanglement)
\cite{Giarmatzi2021Apr, Taranto2024May}.
In the spirit of this
distinction, we here investigate how  memory effects depend on the
measurement instruments, using the paradigmatic L\"uders (preserving quantum memory) and von-Neumann (preserving \textit{no}
quantum memory) measurements as representative types of instruments.

We conclude this discussion with a relevant observation. The repeated
measurement of a Markovian stochastic process can recover
non-Markovian statistics \cite{Diosi2008Feb, Diosi2014Nov,
  Megier2020Dec}, even if the measurement is not affecting the
dynamics \cite{David}. This happens because of the loss of information
due to the 
coarse-graining underlying the measurements or \textit{lumping}
\cite{Blom_2024,holography,Zhao,Tassilo}. To focus on this phenomenon,
we will in the following assume that the dynamics between quantum
measurements is fully Markovian, i.e.\ no memory can be stored in the
environment. We consider that the measurement statistics are Markovian
when they satisfy Eq.~\eqref{eq:markodefinicion} given a fixed
measurement instrument \cite{Milz2020Dec, Smirne2011Sep,
  Smirne2018Nov}.


We now consider Eq.~\eqref{eq:prob_definicion} explicitly for ideal
von-Neumann and L\"uders measurements, respectively, and discuss their connection to
Eq.~\eqref{eq:markodefinicion}.
For simplicity we restrict ourselves to
the scenario of \emph{stroboscopic measurements} (see
Fig.~\ref{fig:esquema}d). In this scenario the measurements are taken
at fixed equidistant times $\tau_n =n\tau$ 
for $n\in \mathbb{N}_0$. Since the environment has no memory, we assume that $\Lambda(t_2, t_1)$ is a completely positive trace-preserving map depending on $\tau$ exclusively, $\Lambda(t_2, t_1)\equiv\Lambda(\tau = t_2  - t_1) = \Lambda_\tau$. After $n+1$ ideal consecutive measurements we obtain the joint probability distribution of the outcomes \cite{gardiner2004quantum} 
\begin{align} 
    &p_n[\{x_i,\tau_i\}_{i\in [0,n]}] =\nonumber\\ &\quad \quad \Tr{\Pi_{x_n}\!\circ\! \Lambda_\tau\! \circ\! \Pi_{x_{n-1}}
      \!\circ \! \Lambda_\tau \circ\!\cdots\!\Pi_{x_0}\!\circ\!\rho_0},
    \label{eq:distribution}
\end{align}
where ${\Pi_x\circ \rho \equiv \Pi_x\rho\Pi_x}$, and ${ \Lambda_\tau\circ \rho \equiv  \Lambda_\tau\rho
   \Lambda_\tau^\dag}$.\color{black}

Interestingly, the distribution in Eq.~\eqref{eq:distribution} in general does \emph{not} satisfy the Kolmogorov condition $p_{n-1}[\{x_i,\tau_i\}_{i\in [0,n]\setminus k}] =
\sum_{x_k}p_n[\{x_i,\tau_i\}_{i\in [0,n]}]$ for some ${k \in [0,n]}$, because the measurement destroys coherence \cite{Breuer2016Apr} and some multi-time correlations may be ill defined \cite{Smirne2018Nov}. Nevertheless, we can always consider the conditional probability of consecutive
\emph{ideal} measurements $p[x, \tau_n|
  \{x_i,\tau_i\}_{i\in [0,n-1]}] \equiv p_n[\{x_i,\tau_i\}_{i\in
    [0,n]}] / p_{n-1}[\{x_i,\tau_i\}_{i\in [0,n-1]}]$
\cite{Smirne2018Nov} and using Eq.~\eqref{eq:distribution} we obtain
\begin{align} \label{eq:transition_matrix}
    p[x_n, \tau_n|  \{x_i,\tau_i\}_{i\in [0,n-1]}] &= p[x_n, \tau_n|
      x_{n-1}, \tau_{n-1}] \nonumber \\&\equiv T_{x_n}^{x_{n-1}},
\end{align}
where $T_x^y =  \left|\bra{x} \Lambda_\tau\ket{y}\right|^2$ is the transition
matrix, whose existence implies that \emph{ideal} outcomes are
Markovian \cite{breuer2002theory}. We now turn to imperfect measurements. Consider first the classical lumping. We set the shorthand notation
${p_n^C[\{X_i,\tau_i\}]\equiv p_n^C[\{X_i,\tau_i\}_{i\in [0,n]}]}$
such that
\begin{align}
p_n^C[\{X_i,\tau_i\}]&=\left (\prod_{k=0}^n\sum_{x_k\in X_k}\right)p_n[\{x_i,\tau_i\}] \\
&= \left (\prod_{k=0}^n\sum_{x_k\in X_k}\right )T^{x_{n-1}}_{x_n}
p_{n-1}[\{x_i,\tau_i\}],\nonumber
\label{classical_lumping}
\end{align}  
where we used Eq.~\eqref{eq:transition_matrix}.~Note that if (and only if)
\begin{equation}
 \!\! \sum_{x_k\in {X_k}} \!\!\!T^{x_{k-1}}_{x_k} \!\!= \!\!\!\sum_{x_k\in {X_k}}\!\!\!
  T^{y_{k-1}}_{x_k},\, \forall x_{k-1},y_{k-1}\in X_{k-1},\,k\in\mathbb{N}
   \label{eq:condition}
\end{equation}
and any mesostate $X_{k-1}=A, B,\ldots$, the classical lumping is
Markovian with transition matrix $\overline{T}_X^{Y}\equiv
\sum_{x\in X}T^{y\in Y}_{x}$
\begin{equation}
  p_n^C[\{X_i,\tau_i\}]=\overline{T}^{X_{n-1}}_{X_n}p_{n-1}^C[\{X_i,\tau_i\}].
    \label{eq:distribution_C}
\end{equation}
Otherwise, the classical lumping is non-Markovian. The condition in
Eq.~\eqref{eq:condition} implies that any mesoscopic transition $A\to
B$ for a given strobosopic time $\tau$
is observed with equal probability, and it is thus
impossible to store information in the history of mesostates
analogously as for
classical dynamics \cite{Lapolla2019Nov}. 

The stroboscopic quantum lumping is analogous to ideal measurements
in Eq.~\eqref{eq:distribution} but with the high-rank projectors
$\oPi_X$ replacing $\Pi_x$, $\oPi_X\circ\rho\equiv\oPi_X\rho\oPi_X$, yielding
\begin{align}
  &p^Q_n[\{X_i,\tau_i\}] =\nonumber \\
  &\quad\quad\Tr{\oPi_{X_n}\! \circ \! \Lambda_\tau\!\circ
   \oPi_{X_{n-1}} \!\circ\!  \Lambda_\tau\! \circ\! \cdots\oPi_{X_0}\!\circ
   \!\rho_0}.\
 \label{eq:distribution_Q}
\end{align}
Denoting by $\delta_{x,y}$ the Kronecker delta we
can partition Eq.~\eqref{eq:distribution_Q}  as ${p^Q_n[\{X_i,\tau_i\}] = p^C_n[\{X_i,\tau_i\}]  + Q_n[\{X_i,\tau_i\}]}$ with
\begin{widetext}
\begin{align}\label{eq:detectable_coh}  
 Q_n[\{X_i,\tau_i\}] =\left(\prod_{k=0}^n\, \sum_{x_k,y_k\in X_k}
  \right)\left(1- \prod_{k=0}^{n-1}\delta_{x_k,y_k}\right)\times
\Tr\{\Pi_{x_n} \Lambda_\tau
   \Pi_{x_{n-1}}  \Lambda_\tau \cdots \rho_0 \cdots  \Lambda_\tau^\dag \Pi_{y_{n-1}} \Lambda_\tau^\dag
      \Pi_{y_n}\},
\end{align}
\end{widetext}
which is the \textit{detectable coherence} that remains after the
respective collapses in the L\"uders measurement
\cite{Smirne2018Nov}. It depends on all the previously observed (meso)states of the
system, including $\rho_0$, so the joint probability distribution in
Eq.~\eqref{eq:distribution_Q} is generally \emph{not} Markovian. The
detectable coherence is a purely quantum memory effect having \emph{no} classical
analogue. If $Q_n = 0$ the classical and quantum lumped statistics coincide
$p^Q_n[\{X_i,\tau_i\}] = p^C_n[\{X_i,\tau_i\}]$ and the coherence is either not generated in the unitary
evolution or cannot be detected by the measurement. 

Therefore, the classical lumped process is
Markovian under the condition in Eq.~\eqref{eq:condition}, whereas
quantum lumping additionally requires ${Q_n = 0}$ for all sequences
$\{X_i,\tau_i\}_{i\in [0,n]}$  for the quantum-lumped process to be Markovian. If either
Eq.~\eqref{eq:condition} is violated or ${Q_n\ne 0}$, the observed process has
memory. This is our main result.

 Remark that, opposite to the von Neumann measurement, the L\"uders
 measurement is independent on how we define the outcomes within the
 mesostates. In fact, any rotation of the outcomes basis which
 preserves the mesostates invariant, gives place to the same quantum
 lumping. The condition on Markovianity of the quantum lumping is independent on how we choose the outcome basis in Eqs.~\eqref{eq:condition} and \eqref{eq:detectable_coh} (see Appendix).

\section{An example} \label{sec:example}

We illustrate the above results with 
an $N$-site quantum lattice (see Fig.~\ref{fig:esquema}d for $N = 4$) with states $\ket{x}$ and Hamiltonian
\begin{equation}
    H = -\sum_{x = 0}^{N - 1} \left(\ket{x+1}\bra{x} +
    \ket{x}\bra{x+1}\right).
    \label{eq:Juan-Milton}
\end{equation}
\color{black} We consider $\hbar = 1$ and periodic boundary conditions \color{black}. Without loss of generality  $N$ is assumed to be even. The
eigenenergies are given by $\epsilon_k = -2\cos(2\pi k/N)$ with
corresponding eigenstates $ \langle x|\epsilon_k\rangle = \e^{-i2\pi k
  x/N}/\sqrt{N}$ for  $k = 0, 1,\ldots, N-1$
\cite{Cohen-Tannoudji2020Jun}. \color{black}We assume a unitary evolution between consecutive measurements $\Lambda_\tau =U(\tau) = \e^{-iH\tau}$. \color{black} We first consider an ideal measurement
detecting the system at site $\ket{x}$ and returning the outcome
$x$. The transition matrix in Eq.~\eqref{eq:transition_matrix} reads
\begin{align}\label{eq:Txy_maintxt}
    T_x^y &= \!\!\left|\bra{x}\e^{-iH\tau}\ket{y}\right|^2 
    \!\!
    \\
    &=\!
    \frac{1}{N^2}\sum^{N-1}_{k,l=0}\e^{-i (\epsilon_k - \epsilon_{l})\tau} \e^{-i\frac{2\pi}{N}(k - l) (x - y)},\nonumber
\end{align}
\begin{figure}[h!]
    \centering
    \includegraphics[width=\linewidth]{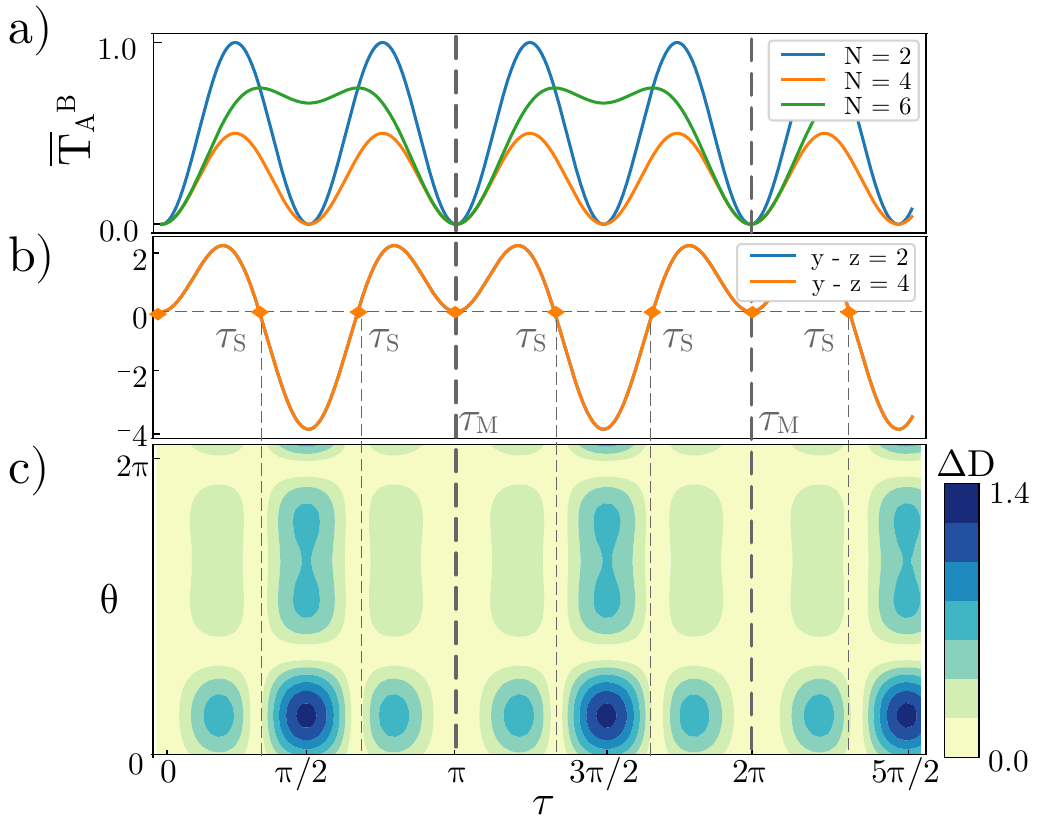}
    \caption{a)~$\overline{T}_A^B$ as a function of $\tau$ for $N = 2,4,6$.~b)
      Sum in Eq.~\eqref{eq:suma} for $N = 6$ and $y-z = 2,4$. The
      crossings with the horizontal line indicate the roots $\tau_{\rm
        M}, \tau_{\rm S}$.~c) $\Delta D_2$ for several values of
      $\tau$ and different initial preparations $\rho_0, \sigma_0$
      parametrized by $\theta$: $\rho_0 =
      \ket{\psi_\theta}\bra{\psi_\theta}$ with
      $\sqrt{2}\ket{\psi_\theta} = \ket{0} + \cos\theta\ket{2} +
      \sin\theta\ket{4}$ and $\sigma_0 = \ket{0}\bra{0}/2 +
      \ket{2}\bra{2}/4+\ket{4}\bra{4}/4$. The quantum lumped process
      is Markovian iff $\Delta D_2 \leq 0$.}
    \label{fig:Dependencia_temporal}
\end{figure}

In considering imperfect observations we define two mesostates $A =
\{2j\}_{j = 0}^{N/2-1}$ and $B = \{2j + 1\}_{j = 0}^{N/2-1}$ (see
Fig.~\ref{fig:esquema}d for $N=4$). This choice is based on real
applications in quantum metrology \cite{Pechal2021Nov} and computation
\cite{Briegel2009Jan}. In the last part of the paper, and also in Appendix \ref{app:boundary chain}, we discuss
that these mesostate configurations appear naturally in qubit chains
under local measurements. We have for any $N\ge 4$
\begin{equation}\label{eq:TAB_example}
    \sum_{x\in A}T_x^y =
    \frac{1}{2} 
    -
    \frac{1}{2N} \sum_{l =0}^{ N-1} \e^{-i 4 \cos\left(\frac{2\pi
        l}{N}\right)\tau}\equiv \overline{T}_A^B,
\end{equation}
and since $\overline{T}_A^B$ does not depend on $y\in B$ the
condition~\eqref{eq:condition} is satisfied and $p^C_n$ is Markovian
for any $\tau$.

In contrast, 
quantum lumping 
is generally non-Markovian
because of the additional non-zero term $Q_n$; for simplicity we here
consider only $n=1$, the result for any $n$ is given in
Appendix~A. According to Eq.~\eqref{eq:distribution_Q} we find,
setting $\rho_0^{yz} \equiv \bra{y}U\rho_0U^\dag\ket{z}$,
\begin{equation}\label{eq:quantum_2}
\begin{split}
    &Q_1[A, \tau;B,0] =\frac{1}{2N}\times
    \\
    &\quad \!\!\!\sum_{z,y\neq z\in B}\!\sum_{l = 0}^{N-1}
   \! \left(1 - \e^{-i 4 \cos\left(\frac{2\pi l}{N}\right)\tau} \right)\e^{i\frac{2\pi}{N}l(y - z)} \rho_0^{yz}.
\end{split}
\end{equation}

$Q_1$ is zero for $A\to B$ and $B\to A$ if (i) the initial state contains no
coherence, $\rho_0^{yz} = 0$,~(ii) if $\e^{-i 4 \cos(2\pi l\tau/N)}=1$, \color{black} or \color{black}(iii) if the sum over $l$ in
Eq.~\eqref{eq:quantum_2} vanishes. (i) depends on the particular
preparation of the system, \color{black}while \color{black}(iii) corresponds to values $\tau = \tau_{\rm S}$ at which 
\begin{equation} \label{eq:suma}
   \!\!\!\! \sum_{l = 0}^{N-1}
    \!\!\left(1 \!-\! \e^{-i 4 \cos\left(\frac{2\pi l}{N}\right)\tau_{\rm S}}
    \right)\!\e^{i\frac{4\pi}{N}l(y-z)}\!\!=\!0,\,\forall z,y\neq z\in B.\
\end{equation}
At stroboscopic times $\tau_S$, the quantum
and classical lumped distributions coincide, $p_n^Q = p_n^C$ and are both
Markovian as they evolve with time. In
Fig.~\ref{fig:Dependencia_temporal}b we show the sum~\eqref{eq:suma}
for $y-z = 2,4$ and $N = 6$ to confirm our prediction. In this
particular case the curves coincide \color{black} and we have \color{black}$\tau_{\rm S} = s\pi/3$.

\color{black}

The case (ii) corresponds to the ``magic'' times $\tau_{\rm M}$ satisfying $4 \cos(2\pi
l\tau_{\rm M}/N) = \pm s2\pi$ for $s\in\mathbb{N}_0$ and $l =
0,\ldots, N-1$. At these times, the transition matrix $\overline{T}_A^B$ vanishes.~Fig.~\ref{fig:Dependencia_temporal}a depicts $\overline{T}_A^B$ for several
values of $\tau$;~the ``magic'' times are $\tau_{\rm M} = s\pi/2$ for
$N = 4$, $\tau_{\rm M} = s\pi$ for $N = 6$, whereas for $N = 8$ we
find $\tau_{\rm M} = 0$ as the unique solution.~Stroboscopic
measurements with $\tau_{\rm M}$ are \emph{not} able to detect any
dynamics between the mesostates even if 
transitions do in fact occur.~A similar phenomenon was observed in first-detection problems~\cite{Yin2019Nov, Thiel2018Jun, Friedman2017Mar}.
\color{black}

We now confirm the presence of purely quantum memory.~We evaluate
the
marginal distributions of the $n$-th outcome, $P^Q_n[X]\equiv(\prod_{k=0}^{n-1}\sum_{X_k})p^Q_n[\{X_i,(i-1)\tau\}]$ with two different initial preparations
$\rho_0$ and $\sigma_0$ parametrized by an angle $\theta$ (see
caption to Fig.~\ref{fig:Dependencia_temporal}).~The Kolmogorov
distance between them is $D_n = \sum_X |P^Q_{n-1}[X](\rho_0) -
P^Q_{n-1}[X](\sigma_0)|$.~If the lumped distributions are Markovian,
the trace distance is nonincreasing between consecutive
measurements, $\Delta D_n = D_{n+1} - D_n \leq 0$ for any $n\in\mathbb{N}$; otherwise the
evolution is non-Markovian~\cite{Breuer2016Apr}.~Fig.~\ref{fig:Dependencia_temporal}c shows ${\Delta D_1 = D_2 - D_1}$
for different values of $\tau$ and different $\rho_0$ and $\sigma_0$. The quantum-lumped process is non-Markovian
in the blue regions where ${\Delta D_1>0}$ and vanishes at $\tau_{\rm S}$
and $\tau_{\rm M}$ in agreement with our prediction. As expected, the
memory also depends on initial correlations, with
unobservable coherences appearing in two horizontal bands with ${\Delta D_1 = 0}$ at two different values of $\theta$, independent of $\tau$.

\section{Discussion}

We analyzed the non-Markovian behavior resulting
from imperfect stroboscopic quantum measurements.  
The von Neumann imperfect measurement produces a mixed quantum state
averaged over all lumped (i.e.\ degenerate) outcomes. Conversely, 
the L\"uders measurement results in a multi-dimensional projection
preserving some quantum coherence.
In both cases, the statistics of the indistinguishable measurement
outcomes are coarse-grained, lumped distributions, which evolve in
time in a generally non-Markovian fashion. The presence and
characteristics of memory were shown to depend tacitly on the measurement, in
particular on the nature of, and projection onto, mesostates.
As a main result, the quantum-lumped process requires stronger
conditions to be Markovian because of preserved quantum
coherence. In particular, it requires the observation to have \emph{no}
detectable coherence. In turn, we introduced the concept of \emph{purely
quantum} memory, which has \emph{no classical counterpart}. It
corresponds to the situation where the respective mesoscopic transition for a given strobosopic time $\tau$ are observed
with equal probability (i.e.\ are Markovian with respect to von
Neumann lumping) and the memory emerges purely due to detectable
coherence. We illustrated these results by means of a quantum walk on
a lattice and determined the parameters for which the respective
lumped distributions are Markovian. 
 The condition Eq.\eqref{eq:condition}, as well as the magic times
 $\tau_M$, are very sensitive to the symmetry of the quantum lattice
 and the mesostates choice.

\begin{figure}[h!]
    \centering
    \includegraphics[width=\linewidth]{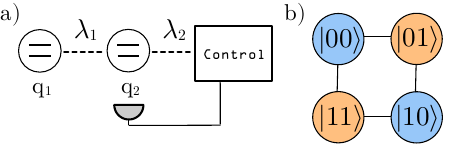}
    \caption{Schematic of the qubit chain setup:~a)
      schematic of the interactions between the qubits and the
      external control. Measurements are performed on $q_2$
      exclusively.~b) Possible transitions between the
      two qubits states created by arbitrary couplings $\lambda_1$ and
      $\lambda_2$. Colors denote the respective indistinguishable
      states under measurements of $q_2$, analogous to those the
      studied in the quantum lattice with $N = 4$.}
    \label{fig:qubit_chain}
\end{figure}

As mentioned, the L\"uders measurement is
noise free whereas the von Neumann version applies in the
presence of strong decoherence, where we note that the relationship between
the stroboscopic measurement time $\tau$ and the decoherence timescale
may be very relevant in practice. 
Future work should extend our results to more realistic measurement
models,  explicitly introducing dissipative (Lindblad) dynamics. A
simple, but conceptually interesting design of this kind is sketched
in Fig.~\ref{fig:qubit_chain}a. The device consists in two coupled
qubits $q_1$ and $q_2$. An external observer measures only $q_2$. In
Appendix B we show that, in the absence of
noise, these measurements behave as L\"uders measurements because of
the coherence remaining in $q_1$. In the presence of a strong
decoherence mechanism, this coherence is removed and von Neumann
measurements are recovered. Weak decoherence mechanisms are in turn
expected to define intermediate types of lumping. The lumpings created
by this setup are equivalent to that studied in the \color{black}article\color{black}, with $N=4$, see Fig.~\ref{fig:qubit_chain}b.

\color{black}
This connects our results to other works such as
Ref.~\cite{Giarmatzi2021Apr}, where non-Markovianity was analyzed from
the perspective of
the distinction between a quantum and a classical environment. In
fact, Fig.~\ref{fig:qubit_chain} shows how a L\"uders imperfect
measurement becomes analog to an ideal measurement on an smaller
system (qubit 2) entangled with an environment (qubit 1). This analogy
will depend on the mesostate configuration and the Hamiltonian
connecting them. In~\cite{Giarmatzi2021Apr} the authors provide a method
to discern between classical and quantum memory, employing
entanglement-detection techniques. The consecutive application of
von-Neumann and L\"uders measurements in turn also provides a simple method to
discern quantum coherence, as shown by our example. These two tests are
in fact complementary and can both be used in a laboratory,
depending on the features and capabilities of the experimental device. 

The above discussion relates to the ongoing debate about how Markovian
behaviour is recovered in quantum systems from pure dynamics (see, for
example, Refs.~\cite{Strasberg2023Jul2,Figueroa-Romero2019Apr,
  Figueroa-Romero2020Sep, Figueroa-Romero2021Jun}).
In fact, given a
multipartite  quantum system under local or imperfect measurements,
the recovered statistics happen to be non-Markovian.
The aforementioned works show that many-body quantum systems recover
Markovianity under certain approximations. In the present work we derived \textit{exact} conditions
required to recover Markovian behavior, where both classical and
quantum memory are absent, independent of the size of the system.
However, we observe that the respective conditions in practice require
extremely symmetric Hamiltonians and mesostate configurations. This sensitivity
of memory effects in small systems can be exploited to improve the
performance of quantum devices, where Markovianity plays an important
role, for example, in quantum thermodynamics (see discussion in
\cite{Zambon2025May}), quantum communication \cite{Cheong2022Nov}, and
control \cite{Sakuldee2018Sep}. However, one of the main motivations
for the present work was to establish a first connection between properties of
quantum measurements and certain important ongoing debates  in the
non-equilibrium thermodynamics of classical systems~\cite{Haenggi_1,Grigolini,Massi_CG,Rahav,Puglisi,Teza,Talkner,martinez,hartichCommentInferringBroken2022,David,Blom_2024,holography,Zhao,Tassilo}. These
works highlight that lumped measurements play a central role in
defining and inferring dissipation (i.e.\ the thermodynamic entropy
production) in coarsely observed systems out of equilibrium. To the
best of our knowledge this debate remains unexplored from a
quantum perspective.

\color{black}

Note that the effect of measurements on (de)coherence in open quantum systems is highly non-trivial. For example, creating
coherence during Landauer's erasure implies additional energy
dissipation \cite{VanVu2022Jan}. Conversely, it is possible to
extract work from coherences during a quantum measurements
\cite{Rio2011Jun, Kammerlander2016Feb}. Other works revealed the
implications of measurement-induced decoherence in entropy generation
\cite{Elouard2017Mar} and recovering thermalization
\cite{Ashida2018Oct}. Our work motivates generalizations of
the above results to imperfect measurements and provokes fundamental questions 
about the implications of quantum/classical lumping for quantum thermodynamics.
An example is the counting statistics of systems under continuous
monitoring, such as electric transport along quantum dots in contact
with electric Fermionic leads or photoelectric detection
\cite{Landi2024Apr,Wiseman1993Jan, Wiseman1994Feb, BibMilburn,  carmichael1993open}. In these experiments the
environment (the electromagnetic field or the Fermionic leads)
'performs' the measurements, because small changes in the quantum
system can be directly detected as single-photon emission, or single-electron transport. This typically corresponds to a von Neumann
lumping, since the only source of noise in the results is generically 
measurement noise. However, the example in Fig.~\ref{fig:qubit_chain}
teaches us that the von-Neumann measurement requires a strong
dephasing during the measurement process. Since the control and
isolation of quantum technologies are improving fast, L\"uders
measurements will become relevant in the near future, and our
formalism could find important applications in these kinds of
experiments.

Finally, while we focused on 
unitary evolution between
consecutive measurements, our results apply to more complex
inter-measurement evolutions, provided that
Eqs.~\eqref{eq:distribution} and~\eqref{eq:distribution_Q} hold. This
is \emph{not} always the case in open quantum systems, where the
contact with an environment may lead to violations of the quantum
regression formula \cite{breuer2002theory, gardiner2004quantum}. The
relationship between non-Markovianity, the quantum regression formula,
and the quantum measurement has recently been studied  extensively~\cite{Smirne2011Sep, Milz2020Dec, Guarnieri2014Aug}. The lumping
described in this work introduces an additional level of complexity;
the role of the contact with the environment on the measurement
statistics yet remains unexplored.

\section*{Acknowledgments} Financial support from the Alexander von Humboldt foundation (postdoctoral fellowship to JTB) as well as the  German Research Foundation (DFG) through the Heisenberg Program  grant GO 2762/4-1 to AG) and European Research Council (ERC) under the European Union’s Horizon Europe research and innovation program (grant agreement No 101086182 to AG) is gratefully acknowledged.

\bibliographystyle{quantum}
\bibliography{biblio}

 \onecolumn\newpage
\appendix
\section{Details on the quantum lattice} Here we derive
equations Eqs.~\eqref{eq:TAB_example} and~\eqref{eq:quantum_2} in the
main text. We start by summing over $x\in A$ in Eq.~\eqref{eq:Txy_maintxt},
\begin{equation}\label{eq:apendice_serie}
    \sum_{x\in A}T_x^y =
    \frac{1}{N^2} \sum_{k,l} \e^{-i (\epsilon_k - \epsilon_{l})\tau} \e^{i\frac{2\pi}{N}(k - l) y} \sum_{x\in A}\e^{-i\frac{2\pi}{N}(k - l)x}.
\end{equation}
This sum depends explicitly on $y$; somewhat surprisingly this
dependence disappears if we restrict $x$ and $y$ to the chosen
mesostates. Observe that the last sum 
\begin{equation}
    \sum_{x\in A}\e^{-i\frac{2\pi}{N}(k - l)x} = \sum_{j = 0}^{N/2 - 1}\e^{-i\frac{4\pi}{N}(k - l)j},
\end{equation}
is a geometric series that we can evaluate to find
\begin{align}
\!\!\!\!\!\!\sum_{j = 0}^{N/2 - 1}\!\!\!\e^{-i\frac{4\pi}{N}(k - l)j} \!=\!
\begin{cases}
	N/2		, & \text{if } k-l = \pm sN/2\\
        \frac{1 - \e^{-i2\pi(k - l) }}{1 - \e^{-i\frac{4\pi}{N}(k - l) }} , & \text{otherwise}\end{cases},
\end{align}  
where $s = 0,1, 2, ...$ is an non-negative integer, $k$ and $l$ are
also integers restricted between $0$ and $N-1$. Since $k-l$ is an
integer the lower 
case \color{black} always gives a null contribution to the series \color{black}and we must only consider $k-l =
\pm sN/2$. In addition, the equations $k-l = \pm sN/2$ have solutions only for $s
= 0,1$: for $s = 0$ the solution is $k = l = 0,1,\ldots, N-1$ while for $s = 1$ we have $k = N/2 + l$ with $l = 0, 1,\ldots, N/2$ and $k = l - N/2$ with $l = N/2,\ldots, N-1$. Inserting this into Eq.~\eqref{eq:apendice_serie} we obtain
\begin{align}
    \sum_{x\in A}T_x^y &=
    \frac{1}{2N} \sum_{k-l = \pm sN/2} \e^{-i (\epsilon_k - \epsilon_{l})\tau} \e^{\pm i\pi s y} \nonumber \\
    &=
    \frac{1}{2} 
    +
    \frac{1}{2N} \e^{\pm i\pi  y} \sum_{k-l = \pm N/2} \e^{-i (\epsilon_k - \epsilon_{l})\tau}.
\end{align}
Since ${y\in B}$ are an odd integers, ${e^{\pm i\pi y} = - 1}$ for all
$B$, so
\begin{equation}
    \overline{T}_A^B =
    \frac{1}{2} 
    -
    \frac{1}{2N} \sum_{k-l = \pm N/2} \e^{-i (\epsilon_k - \epsilon_{l})\tau},
\end{equation}
which satisfies Eq.~\eqref{eq:condition}. Therefore, the classical
lumping process in this particular example is Markovian. To obtain
more insight about the dependence of $\overline{T}_A^B$ on $\tau$ we consider the eigenenergies explicitly,
\begin{align}
    \epsilon_k - \epsilon_l &= -2\cos\frac{2\pi k}{N} + 2 \cos\frac{2\pi l}{N} \nonumber\\
    &= -2\cos\left[\frac{2\pi}{N}( l  \pm N/2)\right] +2 \cos\frac{2\pi l}{N}.
\end{align}
Using the identity ${\cos\left[\frac{2\pi}{N}(l \pm N/2)\right] =-\cos\frac{2\pi l}{N}}$
we have
\begin{equation}
    \epsilon_k - \epsilon_l =  4 \cos\frac{2\pi l}{N},
\end{equation}
from which we obtain Eq.~\eqref{eq:TAB_example} in the main text.

To determine the detectable coherence, we use the spectral
expansion of the Hamiltonian $H$ in Eq.~\eqref{eq:Juan-Milton} to obtain
\begin{align}
    &Q_1[A, \tau;B,0] =\frac{1}{N^2}\sum_{z,y\neq z\in B}\nonumber\\
    & \sum_{k,l} \e^{-i (\epsilon_k -
    \epsilon_{l})\tau+i\frac{2\pi}{N}(ky - lz)} \sum_{x\in
    A}\e^{-i\frac{2\pi}{N}(k - l)x} \rho_0^{yz}.
  \label{analogy}
\end{align} 
The sum over ${x\in A}$ is similar to that in Eq.~\eqref{eq:apendice_serie}, so we consider the restriction ${k-l = \pm sN/2}$, with $s = 0,1$,
\begin{align}\label{eq:tiger_quantumthing}
    &Q_1[A, \tau;B,0] =
    \nonumber\\&=
    \frac{1}{2N} \sum_{z,y\neq z\in B}\sum_{k-l = \pm sn} \e^{-i (\epsilon_k - \epsilon_{l})\tau} \e^{i\frac{2\pi}{N}(ky - lz)} \rho_0^{yz}
    \nonumber\\&=
    \frac{1}{2N} \sum_{z,y\neq z\in B}\sum_{l = 0}^{N-1}\e^{i\frac{2\pi}{N}l(y - z)} \rho_0^{yz} 
   \\&+
    \frac{1}{2N} \sum_{z,y\neq z\in B}\sum_{l = 0}^{N-1} \e^{-i 4 \cos\left(\frac{2\pi l}{N}\right)\tau} \e^{i\frac{2\pi}{N}[l(y-z) \pm \frac{N}{2}y]} \rho_0^{yz}
    \nonumber\\&=
    \frac{1}{2N} \sum_{z,y\neq z\in B}\sum_{l = 0}^{N-1}
    \left(1 - \e^{-i 4 \cos\left(\frac{2\pi l}{N}\right)\tau} \right)\e^{i\frac{2\pi}{N}l(y - z)} \rho_0^{yz}, \nonumber
\end{align}
which is Eq.~\eqref{eq:quantum_2} in the main text.

$Q_n$ for any $n$ follows by noticing that 
${\Tr\{\Pi_{x_n}\cdots\Pi_{y_n}\}=\Tr\{\Pi_{x_n}\cdots\Pi_{x_n}\}\delta_{x_n,y_n}}$
by the cyclic property, writing the trace in
Eq.~\eqref{eq:detectable_coh} as
\begin{align}
& \Tr\{\Pi_{x_n}U
   \Pi_{x_{n-1}} U \cdots \rho_0 \cdots U^\dag \Pi_{y_{n-1}}U^\dag
   \Pi_{y_n}\}=\nonumber\\
   &\quad \delta_{x_n,y_n}\bra{x_{0}}\rho_0\ket{y_{0}}\prod_{i=1}^{n}\bra{x_{i}}U\ket{x_{i-1}}\bra{y_{i-1}}U^\dag\ket{y_i},
\end{align}
and inspecting the $k=n$ term in
Eq.~\eqref{eq:detectable_coh} as
\begin{align}
 &\sum_{x_n\in X_n}
 \braket{x_n|U|x_{n-1}}\braket{y_{n-1}|U^\dag|x_n}  = \frac{1}{N^2}\times\\
 &\sum_{k,l} \e^{-i (\epsilon_k - \epsilon_{l})\tau+i\frac{2\pi}{N}(ky_{n-1} - lx_{n-1})} \sum_{x_n\in X_n}\e^{-i\frac{2\pi}{N}(k - l)x_n}\nonumber.
\end{align}
The above sum is analogous to that in Eq.~\eqref{analogy}, and since it
enters multiplicatively into Eq.~\eqref{eq:detectable_coh}
we conclude that $Q_n$ vanish for the same "magic" times $\tau_M$ and
$\tau_S$ in Eq.\eqref{eq:suma} for any $n$.


\section{Two-qubits example}\label{app:boundary chain}
In most practical applications imperfect measurements occur when we
measure only a controllable part but not a complete quantum system
\cite{breuer2002theory}. The measured portion is known as the
\emph{quantum probe}. The rest of the system remains unaffected by the
measurement, thus
hiding part of the information and in turn giving rise the L\"uders
and von Neumann schemes, as we show. We now consider an explicit example
of a quantum imperfect measurement to gain insight into the respective
measurement schemes that we postulated in the main text. The example
also establishes a link between the example in the main text
and applications in quantum
  technologies. 

We consider a system composed by two qubits, $q_1$ and $q_2$, see
Fig.~\ref{fig:qubit_chain}.~Qubit $q_2$ plays the role of the
quantum probe and can be measured and manipulated externally. $q_1$ is the
hidden part of the system interacting only with $q_2$. We
consider the Hamiltonian $H = \lambda_1 \sigma_x\otimes\sigma_x +
\lambda_2\mathbb{I}\otimes \sigma_x$. The first term represents the
qubit-qubit interaction and the second one an external driving, controlling qubit $q_2$, which represents the quantum probe in this case. $\lambda_1$ and $\lambda_2$ are the intensity of the couplings.
Note that with this choice of Hamiltonian, this system is
equivalent to the quantum lattice in the main text
(Eq.~\eqref{eq:Juan-Milton}) with $N = 4$ states denoted by $\ket{0} =
\ket{00}, \ket{1} = \ket{01},\ket{2} = \ket{10} $ and $\ket{3} =
\ket{11}$. This boundary-driven qubit-chain model is in fact a
paradigmatic example in quantum computing
\cite{nielsen2010quantum}. In recent years it has found even more applications, for example, in quantum transport \cite{Jacob2024Oct, Barra2015Oct} or quantum thermodynamics \cite{Correa2014Feb, Khandelwal2025Feb}. The local single-qubit measurements deeply affect the performance of both, thermal machines and transport systems \cite{Bresque2021Mar, Jacob2024Oct}.

If the initial state of the joint system is pure and the evolution remains unitary, at an arbitrary measurement time $t$ the state will take the form
\begin{equation}
    \ket{\psi_{12}} = \alpha_{00}\ket{00} + \alpha_{01}\ket{01} + \alpha_{10}\ket{10}+\alpha_{11}\ket{11},
\end{equation}
where $\alpha_{ij}$ are time-dependent coefficients. If we measure $q_2$ on this state and obtain the outcome $q_2 = 0$, the resulting state after the measurement is
\begin{equation}
 \left(\mathbb{I}\otimes\ket{0}\bra{0}\right)   \ket{\psi_{12}} = \left(\alpha_{00}\ket{0} +  \alpha_{10}\ket{1}\right) \otimes\ket{0},
\end{equation}
up to normalization. If we measure $q_2=1$, the resulting state is
\begin{equation}
 \left(\mathbb{I}\otimes\ket{1}\bra{1}\right)   \ket{\psi_{12}} = \left(\alpha_{01}\ket{0} +  \alpha_{11}\ket{1}\right) \otimes\ket{1}.
\end{equation}
Observe that this operation is equivalent to the L\"uders measurement with four outcomes $\ket{00}, \ket{01},\ket{10},\ket{11}$ lumped into two indistinguishable mesostates, $A = \{\ket{00}, \ket{10}\}$ and $B = \{\ket{01}, \ket{11}\}$. In this case, measuring $q_2$ only determines the mesostate and \emph{not} the microscopic state of the system. The associated projectors are precisely $\Pi_A =\ket{00}\bra{00}+\ket{10}\bra{10} =  \mathbb{I}\otimes\ket{0}\bra{0}$ and $\Pi_B =\ket{11}\bra{11}+\ket{01}\bra{01} =  \mathbb{I}\otimes\ket{1}\bra{1}$.

The situation is different if the evolution is dissipative. In this case, the system could relax into a diagonal mixed state
\begin{align}
    \rho_{12} &=  r_{00}\ket{00}\bra{00} + r_{01}\ket{01}\bra{01}\nonumber\\ & + r_{10}\ket{10}\bra{10} + r_{11}\ket{11}\bra{11}.
\end{align}
Here $r_{ij}$ are coefficients obeying $r_{00} + r_{10} + r_{01} +r_{11} = 1$. If we measure $q_2$ and obtain the outcome $q_2 = 0$, the resulting diagonal density matrix is
\begin{equation}
    r_{00}\ket{00}\bra{00} + r_{10}\ket{10}\bra{10},
\end{equation}
up to normalization, and 
\begin{equation}
    r_{11}\ket{11}\bra{11} + r_{01}\ket{01}\bra{01}
\end{equation}
if we measure $q_2 = 1$. The resulting matrices are again lumped into
the same mesostates $A$ and $B$. However, opposite to the pure case
above, the quantum coherence is removed from the system, corresponding
to the von Neumann measurement scheme.

\section{General quantum Markov processes}\label{app:Quantum_Mark}
In the main text, we determine the conditions required by the system in order to recover Markovian behaviour, given a measurement basis which define the von Neumann-L\"uders schemes. To complete our study, in this appendix we derive these conditions in a basis-independent way. Our starting point is the definition of a quantum Markovian lumping,
\begin{equation}
 \!\!\!p^Q_n[\{X_i,\tau_i\}] \!=\overline{T}^{X_{n-1}}_{X_n}p^Q_{n-1}[\{X_i,\tau_i\}] \!
\end{equation}
where $\overline{T}^{X_{n-1}}_{X_n}$ is a transition matrix between mesostates. Substituting Eq.\eqref{eq:distribution_Q} into this definition we have
\begin{equation}
\Tr{\oPi_{X_n}  U
   \oPi_{X_{n-1}} 
   \rho_{n-1} \oPi_{X_{n-1}}U^\dag}=\overline{T}^{X_{n-1}}_{X_n} \Tr{
   \oPi_{X_{n-1}} 
   \rho_{n-1} }
\end{equation}
$\rho_{n-1}$ represents the state of the system before the $n-1$ measurement, which contains information about the initial preparation of the system and the previous measurements. We can write this in a more compact way defining the operators $\Phi_{X_{n}}^{X_{n-1}}=\oPi_{X_{n-1}}U^\dag\oPi_{X_n}  U   \oPi_{X_{n-1}}$, then
\begin{equation}
\Tr{ \Phi_{X_{n}}^{X_{n-1}}
   \rho_{n-1} \oPi_{X_{n-1}}}=\overline{T}^{X_{n-1}}_{X_n} \Tr{\rho_{n-1} \oPi_{X_{n-1}}}.
\end{equation}
Where we used the cyclic property of the trace. The eigenbasis of each operator $\Phi_{X_{n}}^{X_{n-1}}$ is defined by $\Phi_{X_{n}}^{X_{n-1}}\ket{\lambda_{X_{n}}^{X_{n-1}}} = \lambda_{X_{n}}^{X_{n-1}}\ket{\lambda_{X_{n}}^{X_{n-1}}}$, $\lambda = 1,2,..., {\rm rank}(\oPi_{X_{n-1}})$. Expanding the trace on this basis we have
\begin{align}
    &\sum_\lambda \lambda_{X_{n}}^{X_{n-1}} \bra{\lambda_{X_{n}}^{X_{n-1}}}\rho_{n-1}\oPi_{X_{n-1}}\ket{\lambda_{X_{n}}^{X_{n-1}}} =\\
    =&\overline{T}^{X_{n-1}}_{X_n}\sum_\lambda \bra{\lambda_{X_{n}}^{X_{n-1}}}\rho_{n-1}\oPi_{X_{n-1}}\ket{\lambda_{X_{n}}^{X_{n-1}}}.\nonumber
\end{align}
Remark that, in general, one could find solutions to this equation for particular configurations of $\rho_0$, $U$ and mesostate configurations. In these particular cases, the evolution of the quantum lumping is Markovian. However, we can find solutions of the last equation independent of $\rho_0$, which are therefore Markovian for any initial condition. In these solutions, the terms inside the sums in the last equation must cancel one by one,
\begin{align}
    &\lambda_{X_{n}}^{X_{n-1}} \bra{\lambda_{X_{n}}^{X_{n-1}}}\rho_{n-1}\oPi_{X_{n-1}}\ket{\lambda_{X_{n}}^{X_{n-1}}} =
    \\
    =&\overline{T}^{X_{n-1}}_{X_n}\bra{\lambda_{X_{n}}^{X_{n-1}}}\rho_{n-1}\oPi_{X_{n-1}}\ket{\lambda_{X_{n}}^{X_{n-1}}}.\nonumber
\end{align}
for any $\lambda$. This means the operator $\Phi_{X_{n}}^{X_{n-1}}$ is fully degenerated, and its unique eigenvalue is, precisely $\overline{T}^{X_{n-1}}_{X_n}$.

To recapitulate, if the quantum lumping is Markovian for any initial preparation, then $\Phi_{X_{n}}^{X_{n-1}}$ (which only depends on the unitary evolution and the mesostates choice, but not in the outcome basis) is fully degenerated, and the transition $\overline{T}^{X_{n-1}}_{X_n}$ matrix coincides with its unique eigenvalue. The last must be true for all the mesostates $X_n$.

Given two degenerated eigenstates $\ket{\lambda_{X_{n}}^{X_{n-1}}}$ and $\ket{\mu_{X_{n}}^{X_{n-1}}}$, with $\lambda \neq \mu$, we have the identity $\bra{\lambda_{X_{n}}^{X_{n-1}}}\Phi_{X_{n}}^{X_{n-1}}\ket{\lambda_{X_{n}}^{X_{n-1}}} = \bra{\mu_{X_{n}}^{X_{n-1}}}\Phi_{X_{n}}^{X_{n-1}}\ket{\mu_{X_{n}}^{X_{n-1}}}$. Expanding the operator we have
\begin{align}
    &\bra{\lambda_{X_{n}}^{X_{n-1}}}
    U^\dag\oPi_{X_n}  U
    \ket{\lambda_{X_{n}}^{X_{n-1}}} =\sum_{x_n\in X_n}|\bra{x_k}U \ket{\lambda_{X_{n}}^{X_{n-1}}}|^2\\
    =& \sum_{x_n\in X_n}|\bra{x_k}U \ket{\mu_{X_{n}}^{X_{n-1}}}|^2 =\bra{\mu_{X_{n}}^{X_{n-1}}}
    U^\dag\oPi_{X_n}  U
    \ket{\mu_{X_{n}}^{X_{n-1}}} \nonumber
\end{align}
Here $\ket{x_n}$ is a basis of the mesostate $X_n$ as in the main text. The former proves that, if the quantum lumping is Markovian for any initial preparation, Eq.\eqref{eq:condition} is satisfied. We notice that if $\Phi_{X_{n}}^{X_{n-1}}$ is degenerated, the basis $\ket{\lambda_{X_{n}}^{X_{n-1}}}$ is not uniquely determined. In fact, any rotation of $\ket{\lambda_{X_{n}}^{X_{n-1}}}$ preserving the mesostate structure still satisfies the equations in this section, each one defining a particular von Neumann measurement.

Expanding the projectors $\oPi_{X_{n-1}} = \sum_\lambda \ket{\lambda_{X_{n}}^{X_{n-1}}}\bra{\lambda_{X_{n}}^{X_{n-1}}} \equiv  \sum_\lambda\oPi_\lambda$ and substituting in Eq.\eqref{eq:distribution_Q} we see that
\begin{equation}
    p^Q_n[\{X_i,\tau_i\}] = \sum_\lambda \tr{\oPi_{X_n}\circ U \circ \oPi_\lambda\circ...\circ\rho_0}.
\end{equation}
As we can see, in the basis $\ket{\lambda_{X_{n}}}$, the quantum lumping coincides with the classical one defined in Eq.\eqref{eq:distribution_C}, so detectable coherence is zero.


\end{document}